# Determination of van der Waals thin film thickness via characteristic X-ray photon yield

Guoshun Luo[1], Shichao Zhao[1], Xinyu Liu[1], Suguo Chen[2], Yue Wang[2], Ping Zhang[1], Zhigang Lu[1], Shaomeng Wang[1], Sunchao Huang[1*], and Yubin Gong[1†]

[1]National Key Laboratory of Science and Technology on Vacuum Electronics, School of Electronic Science and Engineering, University of Electronic Science and Technology of China

[2]Department of Applied Physics, Xian University of Technology, Xian 710048, China

[*] Email: sunchaohuang@uestc.edu.cn; [†]Email: ybgong@uestc.edu.cn

**Abstract:** Characteristic X-ray generation is essential for producing X-ray sources and for investigating light–matter interactions. Here, we demonstrate that characteristic X-ray generation can also be employed to determine the thickness of van der Waals thin films in a non-destructive manner. Specifically, we establish a sixth-order polynomial model that relates film thickness to the characteristic X-ray yield. The model can both predict the X-ray yield of graphite films at arbitrary thicknesses and infer the film thickness from a corresponding characteristic X-ray yield. We validate the model by comparing its predicted X-ray yields for graphite films with those obtained from Monte Carlo simulations. Additionally, we experimentally validated the reliability and feasibility of the proposed method for extracting graphite-film thickness via X-ray photon-yield characterization. Our results provide a non-destructive approach for determining the thickness of van der Waals thin films.

## Introduction

X-ray technology has been widely used in medical imaging, medical treatment, nondestructive detection, and security inspection since the discovery of X-rays by Roentgen in 1895.[1,2] X-ray generation mechanisms include characteristic X-ray generation, Bremsstrahlung radiation, synchrotron radiation, Cherenkov radiation, transition radiation, Smith–Purcell radiation, and parametric X-ray radiation.[3–6] Among these, characteristic X-ray generation is one of the simplest ways to generate X-rays. Characteristic X-rays are produced when energetic free electrons impinge on a target material, ejecting an inner-shell electron from an atom, thereby creating a vacancy. An outer-shell electron then fills this vacancy, releasing energy in the form of an X-ray photon.[7–9] Unlike other X-ray generation mechanisms, characteristic X-ray generation imposes minimal requirements on the properties of the incident electrons. As long as the kinetic energy of the incident electrons exceeds the binding energy of the inner-shell electron, characteristic X-rays can be produced; however, for efficient generation, the incident kinetic energy is recommended to be substantially larger than the binding energy. In this sense, characteristic X-ray sources can be realized in a rather compact manner.[10–13] For example, an 8 keV characteristic X-ray source can be driven by an electron beam with a kinetic energy of 10 keV or 30 keV, whereas other mechanisms, such as synchrotron radiation and parametric X-ray radiation, require GeV-class electron beams to generate an 8 keV photon. Another feature of characteristic X-ray generation is that the output photon energy is fixed by the binding energy of the target material. An advantage of this feature is that it enables the identification of target material composition, a capability widely exploited in materials science. Additionally, a fixed photon energy is required in X-ray diffraction, which is one of the primary applications of characteristic X-ray generation. A disadvantage, however, is that characteristic X-ray generation lacks photon-energy tunability, which limits its use in scenarios that demand tunable photon energies.

While characteristic X-ray generation is already well established for compositional analysis and X-ray diffraction studies, its application to thickness measurement in van der Waals thin films remains largely unexplored. Notably, sample thickness constitutes a critical parameter governing the performance of van der Waals-driven free-electron X-ray sources.[4,14–17] Van der Waals materials represent a unique class of crystalline solids composed of two-dimensional atomic layers—either as monolayers, such as graphene, or as multilayers, like graphite—held together by weak van der Waals forces.[18–20] This family of materials exhibits an extraordinary range of physical properties, including linear energy-momentum dispersion, exceptionally high intrinsic charge carrier mobility, extreme electromagnetic field confinement, quantum Hall effects, van Hove singularities, sensitivity in in-plane and tunneling pressure sensors, gate-tunable plasmons, tunable photon polaritons, as well as superconductivity and strong light-matter interactions at low energies.[21]

Currently, the thickness measurement of van der Waals thin films is commonly performed using localized surface topography microscopy and destructive cross-sectional imaging techniques, with atomic force microscopy being the most established representatives.[14-15] Atomic force microscopy is widely deployed as a benchmark for vertical step-height verification and layer calibration, although its practical throughput is bottlenecked by inherently slow lateral scan rates (typically 0.5–2 Hz). Additionally, continuous tactile scanning scales up the risks of mechanical tip bluntness and induces irreversible structural tearing to delicate exfoliated flakes. Furthermore, atomic force microscopy (AFM) is inherently incapable of measuring the thickness of free-standing van der Waals thin films. Once a film is transferred onto a TEM grid, for instance, it becomes inaccessible for AFM thickness characterization. To circumvent this limitation, one practical approach is to measure the film thickness by AFM prior to transfer. However, this workflow is both time-consuming and practically challenging, as the same flake can be difficult to locate or may even be damaged during the transfer process. On the other hand, cross-sectional transmission electron microscopy combined with convergent-beam electron diffraction[24] or electron energy-loss spectroscopy log-ratio formulations[25] offers sub-nanometer axial accuracy, but mandates tedious, totally destructive specimen preparation via focused ion beam milling or precarious transfer onto specialized grids, rendering real-time, high-throughput industrial quality control completely impractical.[26,27] Optical contrast configurations and spectroscopic ellipsometry enhance screening velocities but routinely confront fatal deficiencies in lateral spatial resolution and depend heavily on substrate-specific complex refractive indices, thereby limiting their generalizability to standalone, substrate-free low-dimensional architectures.[28–32]

Here, we propose a method to determine the thickness of van der Waals thin films—both free-standing and substrate-supported—by exploiting the relationship between characteristic X-ray photon yield and film thickness. We first establish this relationship via Monte Carlo simulations. Based on the simulated data, we develop a sixth-order polynomial model that directly correlates film thickness with the characteristic X-ray yield. This model is then used to predict the X-ray yield for graphite films of arbitrary thicknesses, and conversely, to extract film thickness from experimentally measured yields. The model's accuracy is validated through comparison with independent Monte Carlo simulations. Overall, our findings offer a non-destructive route for thickness characterization of van der Waals thin films.[33–37]

**Results**

When energetic electrons impinge on a van der Waals film, they gradually lose energy through inelastic scattering events. A fraction of this energy transfer is sufficient to ionize the inner-shell electrons of the constituent atoms, creating core-hole states. These holes are subsequently filled by electrons from higher energy levels, and the energy difference is released as characteristic X-ray photons—whose energies are element-specific and thus serve as a fingerprint of the material composition. Depending on the electron kinetic energy and film thickness, two distinct scenarios may occur. As illustrated in Fig. 1a, if the incident electrons possess sufficient energy or the film is sufficiently thin, the electrons can penetrate through the film. Conversely, as shown in Fig. 1b, if the electron energy is insufficient or the film is too thick, the electrons are fully stopped and absorbed within the film. The thickness of the film therefore governs the total interaction volume and, consequently, the number of ionizing events, establishing the physical basis for thickness-dependent characteristic X-ray yield.[38–46]

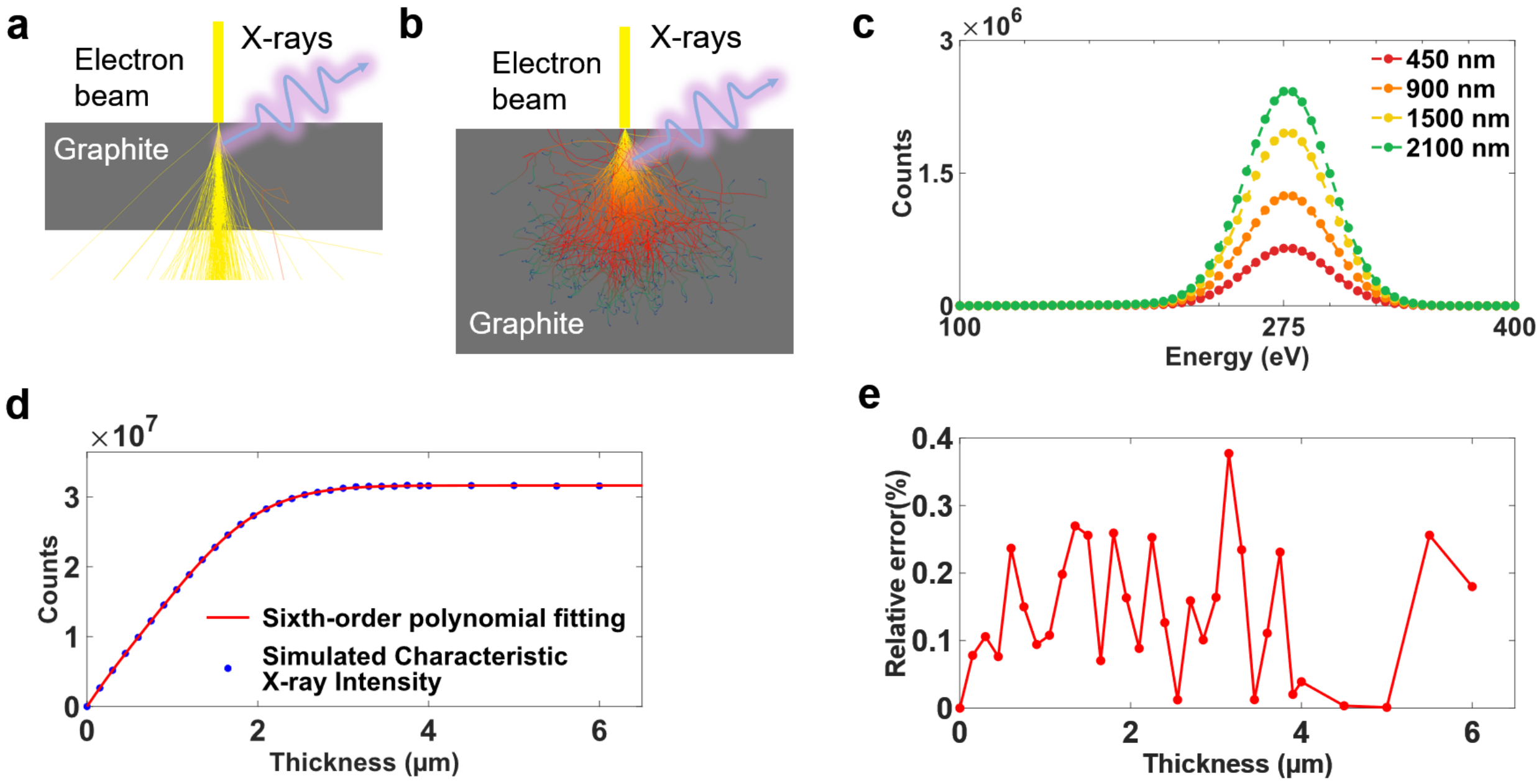

**FIG. 1.** Electron transport and characteristic X-ray yield for freestanding graphite films under 20 keV electron-beam excitation. (a) Simulated electron trajectories in a 200-nm-thick film, for which electrons reach the lower surface. (b) Simulated trajectories in a semi-infinite graphite target. (c) C Kα spectra for film thicknesses of 450, 900, 1500, and 2100 nm. (d) Integrated C Kα yield from the Monte Carlo simulations and the sixth-order model fit. (e) Relative deviation between the fitted and simulated yields; the dashed line marks the 0.4% tolerance.

Figure 1c presents the simulated X-ray spectra obtained from 20 keV electrons impinging on van der Waals thin films with thicknesses of 450 nm, 900 nm, 1500 nm, and 2100 nm. A peak is observed at 277 eV, corresponding to the characteristic K-edge emission of carbon. The peak intensity increases monotonically with film thickness.[47–49]

Figure 1d depicts the dependence of the characteristic X-ray yield on film thickness. In the thin-film regime, the yield exhibits a nearly linear increase with thickness. This behavior arises because the X-ray yield scales with the interaction path length of the incident electrons within the film: the probability of inner-shell ionization increases with the number of scattering events along the electron trajectory. Under the thin-film approximation—where electron energy loss and X-ray self-absorption are negligible—the yield is approximately proportional to the electron path length. Consequently, a path length of $2\Delta l$ yields roughly twice the X-ray intensity of $\Delta l$.[50–53]

This linear relationship, however, holds only when the electron energy remains well above the ionization threshold and the film is sufficiently thin to avoid significant attenuation of the emitted X-rays. As the film thickness increases, this linearity gradually deteriorates due to three combined factors: progressive electron energy degradation, deflection of electron trajectories, and enhanced self-absorption of the generated X-rays. These effects collectively reduce the detected X-ray yield. When the film thickness exceeds the electron penetration depth—defined as the maximum distance that incident electrons can travel within the material—the characteristic X-ray yield reaches saturatio.[7–9]

To characterize this nonlinear dependence, we establish a sixth-order polynomial model combined with a Gaussian damping term to correlate thin-film thickness with characteristic X-ray yield, which is formulated as follows:

$$I(t) = A \cdot \int_0^t \left( \sum_{n=0}^{6} \beta_n z^n \right) \exp(-kz - \left(\frac{z}{R_d}\right)^2) dz \qquad (1)$$

where $I(t)$ designates the absolute generated intensity of characteristic X-rays emitted from a thin-film with thickness $t$ ; $A$ represents the global scaling amplitude and instrument normalization constant capturing primary probe current, collection solid angle, and detector efficiency; $z$ is the independent mass depth variable describing the localization of inner-shell ionization probabilities; $\beta_n$ denote the coefficients of the sixth-order polynomial ($\beta_0$ is fixed to unity, while all other $\beta_n$ serve as fitting parameters); $k$ represents the micrometer-scale photon absorption coefficient defined by $k = \chi\rho/10000$ (where $\chi = \left(\frac{\mu}{\rho}\right)\csc\theta$ , $\frac{\mu}{\rho}$ is the mass attenuation coefficient, and $\theta$ is the observation angle); and $R_d$ denotes the analytical Gaussian truncation parameter which corresponds to the theoretical effective electron excitation range estimated via the classical Kanaya-Okayama (K-O) formulation.[9] Eq. (1) intrinsically captures both the thickness-dependent ionization volume and the attenuation effect, thereby providing the comprehensive theoretical foundation for the non-linear thickness-to-yield inversion mapping established in this study.

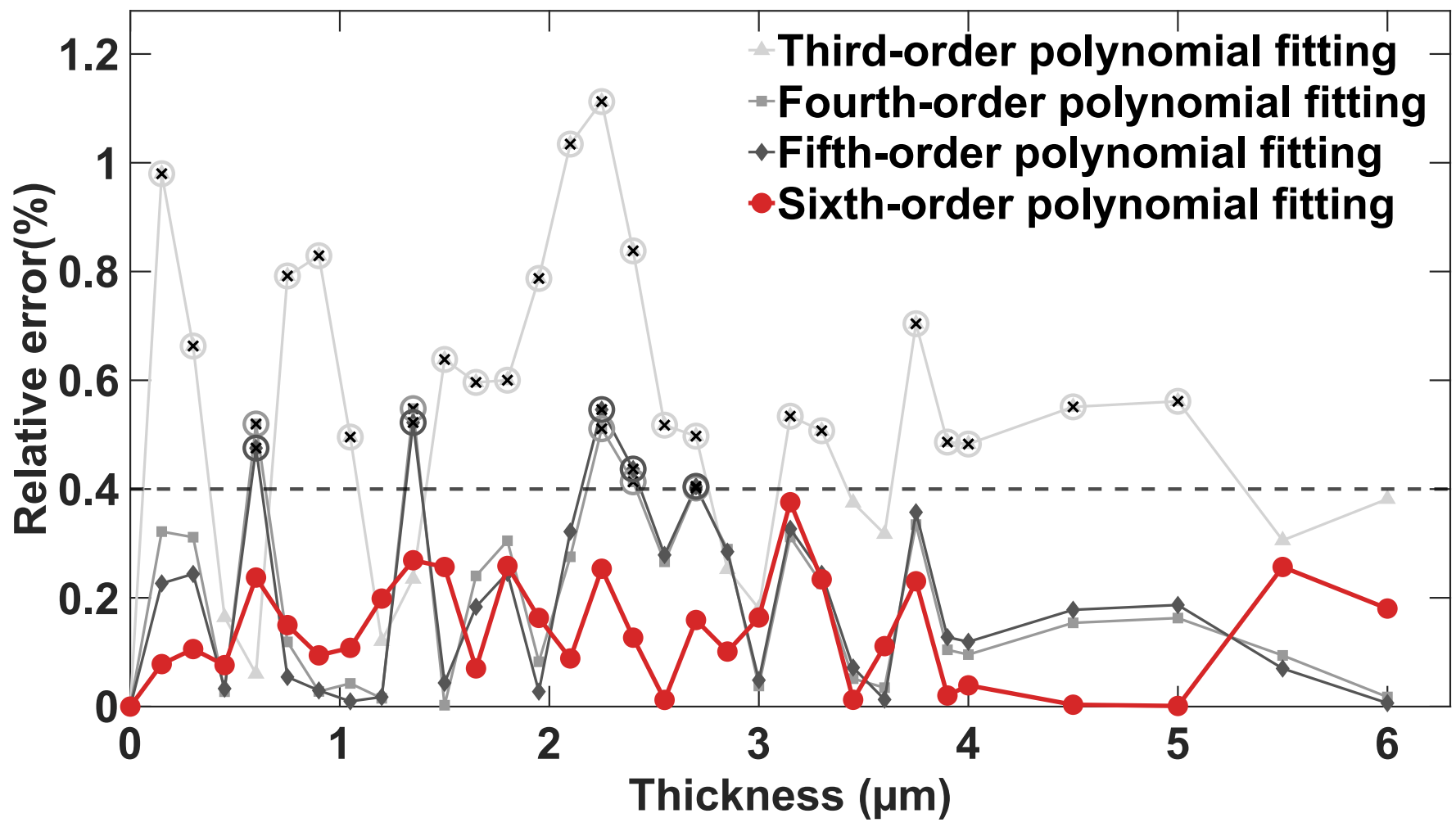

**FIG. 2.** Relative deviations between fitted and simulated X-ray yields for third- through sixth-order polynomial models at an accelerating voltage of 20 kV. The third-, fourth-, and fifth-order results are shown in gray, and the sixth-order results are shown in red. The dashed line marks the 0.4% tolerance, and crossed symbols identify data points above this value.

Figure 2 compares the performance of third-order fourth-order fifth-order and sixth-order polynomial model in describe the relationship of thin-film thickness with characteristic X-ray yield, where electron energy is 20 kV. The maximum relative deviation is 1.1125% for the third-order polynomial. The fourth- and fifth-order models reduce this value to 0.5479% and 0.5461%, respectively. The maximum relative deviation of the sixth-order model is 0.3757%, and all data points are within the 0.4%. The sixth-order model is therefore used in the following analysis.

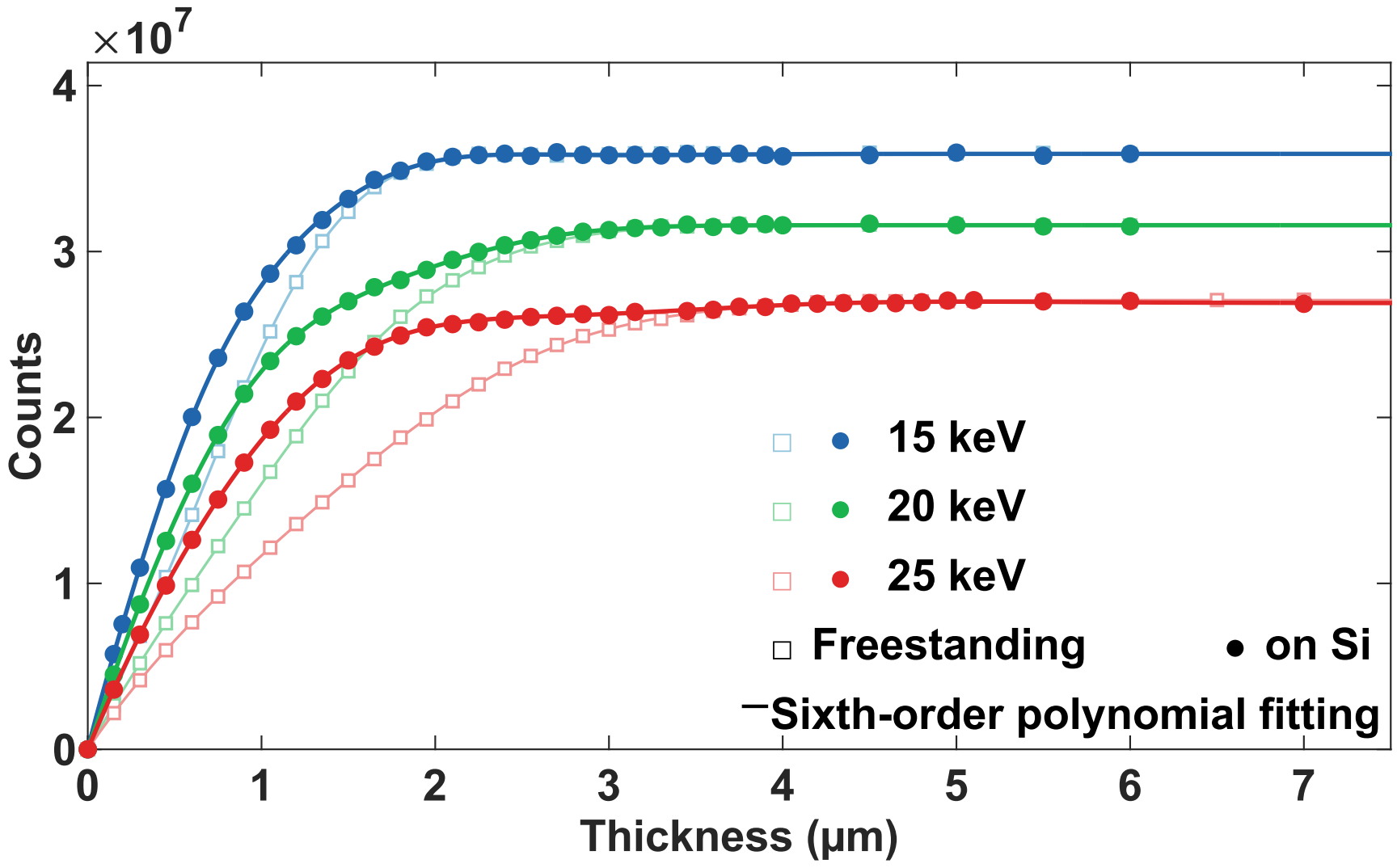

**FIG. 3.** Characteristic X-ray yield as a function of graphite-film thickness for freestanding and Si-supported films. Monte Carlo results and sixth-order model fits are shown at accelerating voltages of 15, 20, and 25 kV. Filled circles denote Si-supported films, and open squares denote freestanding films. At a given voltage, the two curves differ in the thin-film region and converge as the film thickness exceeds the effective electron range.

Figure 3 shows the characteristic X-ray yield as a function of graphite-film thickness for freestanding and $SiO_2$-supported graphite films at three incident-electron energies. Filled circles represent Monte-Carlo simulation data for $SiO_2$-supported graphite, open squares denote results for freestanding graphite, and solid lines correspond to sixth-order polynomial fits. Our results demonstrate that this sixth-order model adequately describes the nonlinear thickness-dependence of the characteristic X-ray yield for both film configurations across the investigated electron energies; the associated fitting parameters are summarized in Table 1.

In the thin-film regime, the $SiO_2$-supported graphite exhibits a higher X-ray photon yield compared with its freestanding counterpart, while the yields converge for sufficiently thick films at all studied energies. This enhancement originates from electrons backscattered at the graphite/$SiO_2$ interface, which re-enter the graphite layer and trigger additional atomic ionization[54–56]. Once the graphite thickness exceeds the effective electron range, most incident electrons dissipate their energy before reaching the substrate interface. The substrate-mediated backscattering contribution thus becomes negligible, and the supported and freestanding films yield nearly identical X-ray intensities.

A clear saturation behaviour is observed in Figure 3. Above the saturation thickness, further increasing the graphite thickness no longer raises the X-ray photon count. This implies that X-ray-yield-based thickness determination cannot be performed for films thicker than this saturation limit. As expected, higher-energy electrons yield larger saturation thicknesses owing to their longer penetration ranges.

**Table 1.** Fitted parameters of the sixth-order model in Eq. (1) for freestanding and $SiO_2$-supported graphite films at accelerating voltages of 15, 20, and 25 kV.

| $E_0$ (keV) | Film configuration | $A$ | $\beta_1$ | $\beta_2$ | $\beta_3$ | $\beta_4$ | $\beta_5$ | $\beta_6$ | $R_d$ (μm) |
|---|---|---|---|---|---|---|---|---|---|
| 15 | Freestanding | $2.30 \times 10^7$ | 0.8604 | 0.8922 | 5.2039 | −4.3325 | 0.8393 | −0.0071 | 1.5085 |
| 15 | $SiO_2$-supported | $3.80 \times 10^7$ | 1.8998 | −4.3057 | 7.1598 | −4.8144 | 1.4954 | −0.1922 | 1.5273 |
| 20 | Freestanding | $1.81 \times 10^7$ | 0.8015 | 0.1966 | 1.2825 | 0.8368 | −0.4925 | 0.1476 | 1.4833 |
| 20 | $SiO_2$-supported | $2.41 \times 10^7$ | 2.1974 | −8.1066 | 16.4075 | −12.7404 | 4.0263 | −0.2892 | 1.3647 |
| 25 | Freestanding | $1.74 \times 10^7$ | 0.1543 | 1.5112 | −2.0125 | 2.0795 | −0.6160 | −0.0649 | 2.2446 |
| 25 | $SiO_2$-supported | $2.63 \times 10^7$ | 0.3540 | 0.5085 | 0.0751 | −0.4761 | 0.2080 | −0.7239 | 2.9500 |

**Note.** $A$, $\beta_1$–$\beta_6$, and $R_d$ are condition-specific fitted parameters. The zeroth-order coefficient $\beta_0$ is fixed to unity and is therefore not listed. When z is expressed in μm, $R_d$ is in μm, $\beta_n$ has units of $\mu m^{-n}$, and $A$ carries the corresponding intensity-normalization unit (counts $\mu m^{-1}$ when I(t) is reported as counts). The attenuation factor is calculated from $(\mu/\rho)\csc\theta$, where the graphite mass attenuation coefficient is $\mu/\rho = 2373\ cm^2\ g^{-1}$ and the detector take-off angle is $\theta = 30°$. Thus, $(\mu/\rho)\csc\theta = 2373 \times 2 = 4746\ cm^2\ g^{-1}$. For the physical-depth form used in $\exp(-kz)$ with z in μm, the equivalent attenuation coefficient is $k = [(\mu/\rho)\rho\csc\theta]/10^4 = 1.0726\ \mu m^{-1}$, using $\rho = 2.26\ g\ cm^{-3}$.

An apparently counter-intuitive trend is also observed: higher-energy incident electrons produce lower saturated X-ray yields than lower-energy electrons. The underlying physical origin lies in the energy dependence of the carbon K-shell ionization cross-section. For C K-α emission, the incident energies of 15–25 keV lie far above the carbon K-shell ionization threshold (~284 eV). In this energy regime, the K-shell ionization cross-section decreases with increasing electron energy. Accordingly, the saturated characteristic X-ray yield decreases at higher accelerating voltages, despite the concurrent increase in electron penetration depth.

**Table 2.** Comparison of sixth-order model predictions and Monte Carlo carbon characteristic X-ray yields for freestanding graphite films at 20 kV.

| *Thickness (nm)* | *Model-predicted X-ray counts* | *Monte Carlo X-ray counts* | *Relative error* |
|---|---|---|---|
| 20 | 361941 | 361689 | 0.07% |
| 60 | 1079224 | 1079409 | 0.02% |
| 150 | 2658275 | 2656590 | 0.06% |
| 300 | 5181156 | 5186927 | 0.11% |
| 600 | 9926577 | 9902943 | 0.24% |
| 1200 | 18904154 | 18867594 | 0.19% |
| 2100 | 28292941 | 28267254 | 0.09% |

**Table 3.** Thicknesses reconstructed from carbon characteristic X-ray yields for freestanding graphite films at 20 kV.

| *X-ray yield (counts)* | *Reconstructed thickness (nm)* | *Reference thickness (nm)* | *Relative error* |
|---|---|---|---|
| 722018 | 40.02 | 40 | 0.04% |
| 1435524 | 80.06 | 80 | 0.08% |
| 3510619 | 199.80 | 200 | 0.10% |
| 7595649 | 450.36 | 450 | 0.08% |
| 14516457 | 900.98 | 900 | 0.11% |
| 18867594 | 1197.41 | 1200 | 0.22% |
| 22778969 | 1495.08 | 1500 | 0.33% |

Without loss of generality, we validate the performance of the sixth-order model (Eq. 1) for describing the nonlinear dependence of X-ray photon yield on graphite-film thickness using 20 keV incident electrons. Two complementary tests are performed to assess model performance. In the first test, the sixth-order model is employed to predict X-ray photon yields across a range of graphite-film thicknesses, and these predictions are compared against reference results from Monte Carlo simulations. Table 2 lists the model-calculated yields alongside Monte Carlo values for film thicknesses spanning 20 nm to 2100 nm. The resulting relative errors fall within 0.02%–0.24%, and remain below 0.4% for all considered thickness points. In the second test, we evaluate the model's capability to retrieve graphite-film thickness from measured X-ray photon yields. We first simulate X-ray yields for a set of known reference graphite-film thicknesses. These yield values are then fed into the sixth-order model to reconstruct the corresponding film thickness. The reconstruction outcomes are compiled in Table 3. For the 20 kV freestanding configuration, the relative deviation between reconstructed and reference thicknesses ranges from 0.04% to 0.33%.

We further validate the feasibility of graphite-film thickness determination based on experimentally measured X-ray photon yields. A set of graphite thin-film samples were prepared via mechanical exfoliation on silicon substrates, and their actual thicknesses were precisely characterized using atomic force microscopy (AFM), with a typical AFM morphology displayed in Fig. 4a. The X-ray photon yields of these samples were experimentally collected using a scanning electron microscope (SEM) integrated with an energy-dispersive X-ray spectroscopy (EDS) detector, and a representative EDS energy spectrum is illustrated in Fig. 4b.

Figure 4c presents the correlation between X-ray photon yield and graphite film thickness. In this plot, green circles correspond to experimental measurements, blue circles denote Monte Carlo simulation results, and the solid line represents the fitted sixth-order polynomial model. To eliminate systematic experimental errors, a correction factor $\beta$ is introduced to calibrate the measured X-ray yields, accounting for the influences of detector collection solid angle and detection efficiency. Figure 4d provides a magnified view of the 0–300 nm thickness region extracted from Fig. 4c for detailed observation. Excellent consistency is observed between experimental measurements and Monte Carlo simulation results, fully demonstrating the reliability and feasibility of the proposed method for determining graphite film thickness using X-ray photon yield characterization.

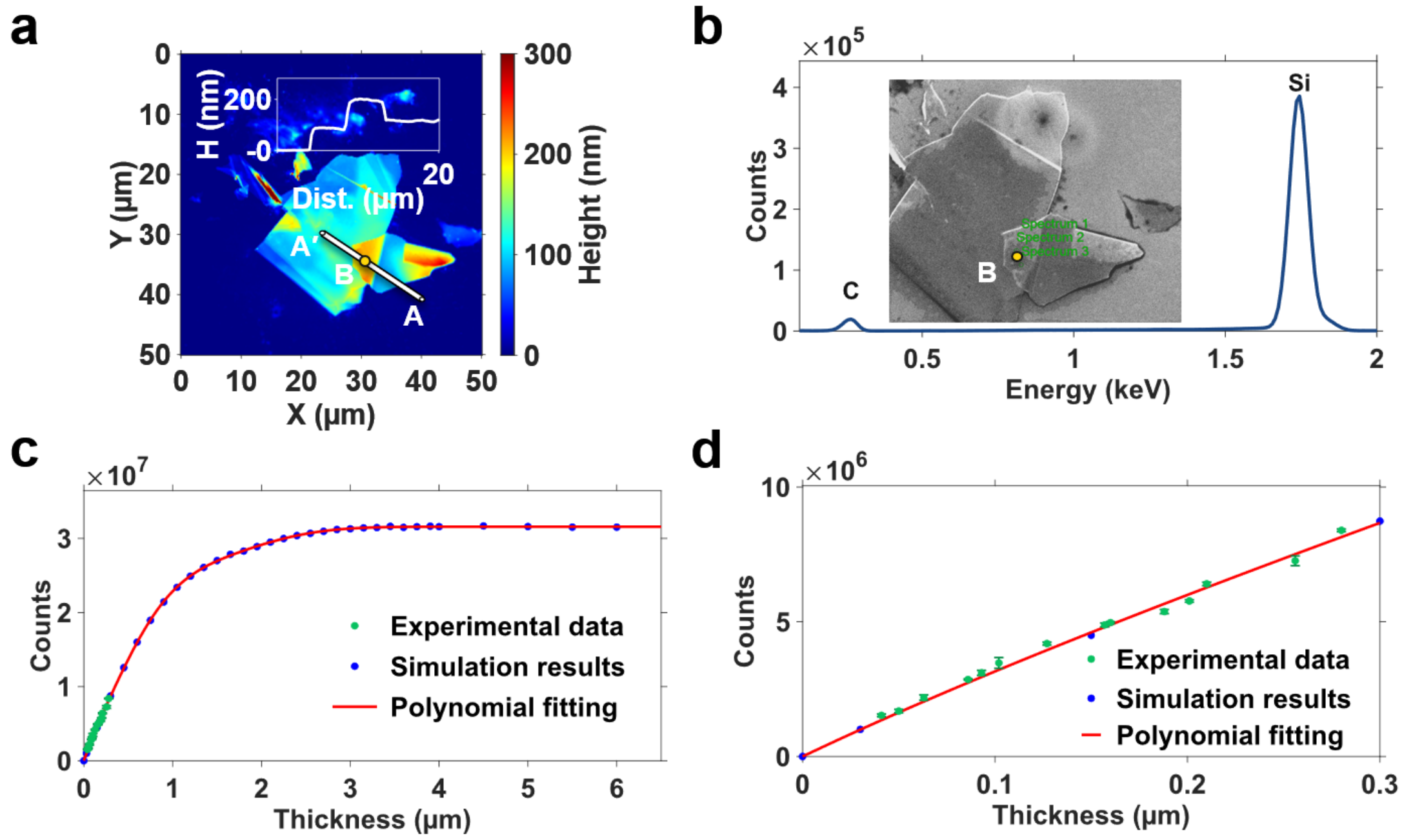

FIG. 4. Experimental validation of the reliability and feasibility of our proposed approach for extracting graphite-film thickness via X-ray photon-yield characterization. (a) AFM image of the graphite sample on a Si substrate. The inset presents the height profile along the marked line AA′. Point B marks the position chosen for EDS measurement, where the graphite thickness derived from the AFM height profile is 201 nm. (b) EDS spectrum collected at point B; the inset displays the corresponding SEM micrograph of the graphite sample with the location of point B indicated. The SEM-EDS measurement was carried out at an accelerating voltage of 20 kV and a beam current of 1 nA. Here, a scaling factor of $\beta = 1.13$ is applied to the experimental results to account for the effects of collection solid angle and detector efficiency. (c) Comparison of experimental measured, Monte-Carlo-simulated, and polynomial-model-predicted X-ray photon yield across the full thickness range. (d) Zoom-in view of panel (c) covering the thickness interval 0–300 nm.

In summary, we have demonstrated that characteristic X-ray generation enables non-destructive thickness determination of van der Waals thin films. Specifically, we developed a sixth-order polynomial model correlating film thickness with characteristic X-ray yield. This model can predict X-ray yields of graphite films for arbitrary thicknesses and, conversely, infer film thickness from measured characteristic X-ray yields. We validated the model by comparing its predicted yields against results from Monte-Carlo simulations. Furthermore, we experimentally verified the reliability and feasibility of our proposed approach for extracting graphite-film thickness via X-ray photon-yield characterization. Collectively, these results offer a viable non-destructive route for thickness metrology of van der Waals thin-film systems.

**Acknowledgement**

This work was supported by National natural science foundation of China (Grants No.62527804, 62588201）

**Data Availability**

The data that support the findings of this study are available from the corresponding author upon reasonable request.

## References

[1] L.J. Wong, and I. Kaminer, “Prospects in x-ray science emerging from quantum optics and nanomaterials,” Appl. Phys. Lett. **119**(13), 130502 (2021).

[2] S. Chen, P. Wang, X. Zhang, S. Huang, Y. Wang, M. Hu, C. Zhang, and Y. Gong, “Thermionic enhanced heat transfer in x-ray tubes,” Appl. Phys. Lett., (2024).

[3] L. Qian, H. Min, Z. Ping, and W. Shao, “Enhancing Heat Transfer in X-ray Tube by van der Waals Heterostructures-based Thermionic Emission,” ACUUM ELECTRONICS, (2026).

[4] S. Huang, R. Duan, N. Pramanik, M. Go, C. Boothroyd, Z. Liu, and L.J. Wong, “Multicolor x-rays from free electron–driven van der Waals heterostructures,” Sci. Adv. **9**(48), eadj8584 (2023).

[5] X. Shi, M. Shentcis, Y. Kurman, L.J. Wong, F.J. García De Abajo, and I. Kaminer, “Free-electron-driven X-ray caustics from strained van der Waals materials,” Optica **10**(2), 292 (2023).

[6] R. Chen, J. Chen, Z. Gong, X. Zhang, X. Zhu, Y. Yang, I. Kaminer, H. Chen, B. Zhang, and X. Lin, “Free-electron Brewster-transition radiation,” Sci. Adv. **9**(32), eadh8098 (2023).

[7] C.J. Powell, “Cross sections for ionization of inner-shell electrons by electrons,” Rev. Mod. Phys. **48**(1), 33–47 (1976).

[8] H. Bethe, “Zur Theorie des Durchgangs schneller Korpuskularstrahlen durch Materie,” Annalen der Physik **397**(3), 325–400 (1930).

[9] K Kanaya and S Okayama, “Penetration and energy-loss theory of electrons in solid targets,” J. Phys. D: Appl. Phys. **5**(1), 43–58 (1972).

[10] M. Vassholz, and T. Salditt, “Observation of electron-induced characteristic x-ray and bremsstrahlung radiation from a waveguide cavity,” Sci. Adv. **7**(4), eabd5677 (2021).

[11] Q. Zhai, N. Pramanik, R. Duan, S. Huang, Z. Liu, and L.J. Wong, “Enhanced tunable X-rays from bulk crystals driven by table-top free electron energies,” Nat Commun **16**(1), 11218 (2025).

[12] W. Eschen, R. Klas, D.S.P. Molina, S. Fuchs, G.G. Paulus, J. Limpert, and J. Rothhardt, “Coherent nanoscale imaging and chemical mapping with compact extreme ultraviolet and soft x-ray sources: Review and perspective,” APL Photonics **10**(5), 050901 (2025).

[13] Y. Liu, L. Li, Y. Liu, X. Zhang, L. Xin, Z. Fu, J. Tian, W. Zhao, and D. Luo, “Optimization of Tungsten Anode Target Design for High-Energy Microfocus X-Ray Sources via Geant4 Monte Carlo Simulation,” Photonics **12**(11), 1062 (2025).

[14] X. Shi, L.W.W. Wong, S. Huang, L.J. Wong, and I. Kaminer, “Transverse recoil imprinted on free-electron radiation,” Nat. Commun. **15**(1), 7803 (2024).

[15] N. Pramanik, S. Huang, R. Duan, Q. Zhai, M. Go, C. Boothroyd, Z. Liu, and L.J. Wong, "Fundamental scaling laws of water-window X-rays from free-electron-driven van der Waals structures," Nat. Photonics **18**(11), 1203–1211 (2024).

[16] S. Huang, R. Duan, N. Pramanik, J.S. Herrin, C. Boothroyd, Z. Liu, and L.J. Wong, "Quantum recoil in free-electron interactions with atomic lattices," Nat. Photonics **17**(3), 224–230 (2023).

[17] S. Huang, R. Duan, N. Pramanik, C. Boothroyd, Z. Liu, and L.J. Wong, "Enhanced Versatility of Table-Top X-Rays from Van der Waals Structures," Adv. Sci. **9**(16), 2105401 (2022).

[18] C. Backes et al, "Production and processing of graphene and related materials," 2D Mater. **7**(2), 022001 (2020).

[19] F. Xia, H. Wang, D. Xiao, M. Dubey, and A. Ramasubramaniam, "Two-dimensional material nanophotonics," Nature Photon **8**(12), 899–907 (2014).

[20] P.G. Zotev, P. Bouteyre, Y. Wang, S.A. Randerson, X. Hu, L. Sortino, Y. Wang, T. Shegai, S.-H. Gong, A. Tittl, I. Aharonovich, and A.I. Tartakovskii, "Nanophotonics with multilayer van der Waals materials," Nat. Photon. **19**(8), 788–802 (2025).

[21] A.J. Moilanen, M. Cavigelli, T. Taniguchi, K. Watanabe, and L. Novotny, "Electrical Control of Photoluminescence in 2D Semiconductors Coupled to Plasmonic Lattices," ACS Nano **19**(4), 4731–4738 (2025).

[22] J.S. Villarrubia, "Algorithms for scanned probe microscope image simulation, surface reconstruction, and tip estimation," J. Res. Natl. Inst. Stand. Technol. **102**(4), 425 (1997).

[23] R. Gröger, T. Heiler, T. Schimmel, and S. Walheim, "Tip-Induced Nanopatterning of Ultrathin Polymer Brushes," Small **19**(29), 2204962 (2023).

[24] S.M. Allen, "Foil thickness measurements from convergent-beam diffraction patterns," Philosophical Magazine A **43**(2), 325–335 (1981).

[25] T. Malis, S.C. Cheng, and R.F. Egerton, "EELS log-ratio technique for specimen-thickness measurement in the TEM," J. Elec. Microsc. Tech. **8**(2), 193–200 (1988).

[26] S. Rubino, S. Akhtar, and K. Leifer, "A Simple Transmission Electron Microscopy Method for Fast Thickness Characterization of Suspended Graphene and Graphite Flakes," Microsc Microanal **22**(1), 250–256 (2016).

[27] M.G. Masteghin, Z.P. Aslam, A.P. Brown, M.J. Whiting, S.K. Clowes, R.P. Webb, and D.C. Cox, "Quantifying the Full Damage Profile of Focused Ion Beams via 4D-STEM Precession Electron Diffraction and PSNR Metrics," Small Methods **10**(5), e02258 (2026).

[28] R. Kenaz, S. Ghosh, P. Ramachandran, K. Watanabe, T. Taniguchi, H. Steinberg, and R. Rapaport, "Thickness Mapping and Layer Number Identification of Exfoliated van der Waals Materials by Fourier Imaging Micro-Ellipsometry," ACS Nano **17**(10), 9188–9196 (2023).

[29] J.N. Hilfiker, N. Singh, T. Tiwald, D. Convey, S.M. Smith, J.H. Baker, and H.G. Tompkins, "Survey of methods to characterize thin absorbing films with Spectroscopic Ellipsometry," Thin Solid Films **516**(22), 7979–7989 (2008).

[30] J. Oh, J. Son, C. Yoon, E. Hwang, J. Ahn, J. Lee, J. Lee, J. Shin, D. Lee, S. Lim, J. Ahn, Y. Sohn, S. Hyun, M. Lee, and T. Jo, "Ultra-wide-field imaging Mueller matrix spectroscopic ellipsometry for semiconductor metrology," Nat Commun **16**(1), 8512 (2025).

[31] P.A. Leger, A. Ramesh, T. Ulloa, and Y. Wu, "Machine learning enabled fast optical identification and characterization of 2D materials," Sci Rep **14**(1), 27808 (2024).

[32] R.S. Sharbidre, P. Narute, J.C. Byen, D. Kim, J. Park, B.C. Park, and S.-G. Hong, "In-Column Backscattered Electron Microscopy for a Rapid Identification of the Number of Layers in $MoS_2$ Nanosheets: Implications for Electronic and Optoelectronic Devices," ACS Appl. Nano Mater. **7**(1), 153–162 (2024).

[33] I. Prencipe, D. Dellasega, A. Zani, D. Rizzo, and M. Passoni, "Energy dispersive x-ray spectroscopy for nanostructured thin film density evaluation," Science and Technology of Advanced Materials **16**(2), 025007 (2015).

[34] T.E. Timofeeva, V.B. Timofeev, V.I. Popov, and S.A. Smagulova, "Estimation of the Thickness of Graphite Nanofilm on a Silicon Substrate by Using Energy Dispersive X-Ray Analysis Data," Nanotechnologies in Russia **11**(7–8), 454–460 (2016).

[35] J.P.N. Cruz, C.M. Garzón, J.K. Noreña, F. Pineda, Juan.S. Cachaya, and A.A.C. Recco, "Accuracy of a reference-free Monte Carlo approach for SEM-EDS thickness assessment of TiN coatings onto diverse substrates," Micron **154**, 103204 (2022).

[36] A. Pazzaglia, A. Maffini, D. Dellasega, A. Lamperti, and M. Passoni, "Reference-free evaluation of thin films mass thickness and composition through energy dispersive X-ray spectroscopy," Materials Characterization **153**, 92–102 (2019).

[37] N.W.M. Ritchie, A. Herzing, and V.P. Oleshko, "Quantification of Unsupported Thin-Film X-ray Spectra Using Bulk Standards," Microscopy and Microanalysis **29**(6), 1921–1930 (2023).

[38] M. Dapor, "Energy loss spectra of low primary energy (E0⩽1keV) electrons backscattered by silicon dioxide," Surface Science **600**(20), 4728–4734 (2006).

[39] H. Gümüş, and Ö. Kabadayi, "Practical calculations of stopping powers for intermediate energy electrons in some elemental solids," Vacuum **85**(2), 245–252 (2010).

[40] F. Salvat-Pujol, H.O. Jeschke, and R. Valenti, "Simulation of electron transport in electron beam induced deposition of nanostructures," Beilstein J. Nanotechnol. **4**, 781–792 (2013).

[41] P. Guo, S.F. Mao, Y.B. Zou, T.F. Yang, H. Miao, and Z.J. Ding, "Monte Carlo simulation study on secondary electron yield of SiO2," Results in Physics **58**, 107472 (2024).

[42] Z.-J. Ding, R. Shimizu, and K. Obori, "Monte Carlo simulation of x-ray spectra in electron probe microanalysis: Comparison of continuum with experiment," Journal of Applied Physics **76**(11), 7180–7187 (1994).

[43] D. Drouin, A.R. Couture, D. Joly, X. Tastet, V. Aimez, and R. Gauvin, "CASINO V2.42—A Fast and Easy-to-use Modeling Tool for Scanning Electron Microscopy and Microanalysis Users," Scanning **29**(3), 92–101 (2007).

[44] R. Gauvin, E. Lifshin, H. Demers, P. Horny, and H. Campbell, "Win X-ray: A New Monte Carlo Program that Computes X-ray Spectra Obtained with a Scanning Electron Microscope," Microsc. Microanal. **12**(01), 49–64 (2006).

[45] D.E. Newbury, and N.W.M. Ritchie, "Simulating electron-excited energy dispersive X-ray spectra with the NIST DTSA-II open-source software platform," MRS Advances **7**(31), 892–898 (2022).

[46] N.W.M. Ritchie, "Spectrum Simulation in DTSA-II," Microsc Microanal **15**(5), 454–468 (2009).

[47] N. Brodusch, and R. Gauvin, "Phase differentiation based on x-ray energy spectrum correlation with an energy dispersive spectrometer (EDS)," Ultramicroscopy **238**, 113534 (2022).

[48] N. Toth, S. Shi, J. Maclennan, and P. Tung, "EDS Analysis for Petrology: A Probabilistic Framework With GPyEDS," Journal of Geophysical Research: Machine Learning and Computation **2**(4), e2025JH000751 (2025).

[49] D. Gao, Y. Yuan, R. Gauvin, and N. Piché, "MC X-ray Coupled with Neural Networks for Element Quantification: A Neural Network-Enhanced Approach," Microscopy and Microanalysis **30**, ozae044.1028 (2024).

[50] J. Donovan, "Multi-Voltage EPMA of Thin Films or Nano Science in the Z Dimension," Microsc Microanal **12**(S02), 878–879 (2006).

[51] J. Fournelle, H. Cathey, P.T. Pinard, and S. Richter, "Low voltage EPMA: experiments on a new frontier in microanalysis - analytical lateral resolution," IOP Conf. Ser.: Mater. Sci. Eng. **109**, 012003 (2016).

[52] T. Schirmer, and T. Ulrich, "A multi-point calibration method for electron probe microanalysis (EPMA) of indium in sphalerite (ZnS)," Sci Rep **15**(1), 7915 (2025).

[53] A. Moy, and J.H. Fournelle, "BadgerFilm: A new thin film analysis programme (with other useful applications for EPMA)," IOP Conf. Ser.: Mater. Sci. Eng. **1324**(1), 012010 (2025).

[54] T. Kowoll, E. Müller, S. Fritsch-Decker, S. Hettler, H. Störmer, C. Weiss, and D. Gerthsen, "Contrast of Backscattered Electron SEM Images of Nanoparticles on Substrates with Complex Structure," Scanning **2017**, 1–12 (2017).

[55] E.I. Rau, and S.V. Zaitsev, "Effect of increasing the coefficient of backscattered electrons for multilayer nanostructures and image contrast inversion in scanning electron microscopy," Phys. Usp. **68**(04), 401–407 (2025).

[56] M. Abbasi, M. Schmid, A. Mafodda, and S. Aschauer, "Enhancing Backscattered Electron Detection in SEM: Investigating Geometric Collection Efficiency and Diode Layout Optimization," Microscopy and Microanalysis **30**, ozae044.1092 (2024).